\documentclass[11pt]{article}
\usepackage[left=1in,top=1in,right=1in,bottom=1in]{geometry}
\usepackage{times}
\usepackage{verbatim}
\usepackage{amsmath}
\usepackage{rotating}
\usepackage{algorithm}
\usepackage[noend]{algpseudocode}

\usepackage{silence}
\usepackage{tabularx}
\usepackage{graphicx}
\usepackage{url}
\usepackage{multirow}
\usepackage{subfig}

\begin{document}
\title{Optimal Weighting for Exam Composition}
\author{Sam Ganzfried and Farzana Yusuf \\
\{sganzfri,fyusu003\}@cis.fiu.edu \\
Florida International University \\
School of Computing and Information Sciences 
%College of Engineering and Computing \\
%11200 S.W. 8th St. Modesto A. Maidique Campus, ECS 381 \\
%Miami, FL 33199
}

\date{\vspace{-5ex}}

\maketitle

\begin{abstract}
A problem faced by many instructors is that of designing exams that accurately assess the abilities of the students. Typically these exams are prepared several days in advance, and generic question scores are used based on rough approximation of the question difficulty and length. For example, for a recent class taught by the author, there were 30 multiple choice questions worth 3 points, 15 true/false with explanation questions worth 4 points, and 5 analytical exercises worth 10 points. We describe a novel framework where algorithms from machine learning are used to modify the exam question weights in order to optimize the exam scores, using the overall class grade as a proxy for a student's true ability. We show that significant error reduction can be obtained by our approach over standard weighting schemes, and we make several new observations regarding the properties of the ``good'' and ``bad'' exam questions that can have impact on the design of improved future evaluation methods.
\end{abstract}

\section{Introduction and Background}                               
\label{se:intro}
Examinations have traditionally been dominant in student performance evaluation, often accompanied with other forms of assessment such as assignments and projects. Defining standards for performance evaluation has been studied from different perspectives~\cite{Norcini03:Setting}. Scouller documented the effectiveness of two different methods---assignment essay vs. multiple choice test---to assess the student ability~\cite{Scouller98:Influence}; in contrast, Kirkpatrick described the negative influences of exam-oriented assessment~\cite{Kirkpatrick11:Negative}. Most relevant question evaluation processes generally emphasize multiple choice testing for measuring students' knowledge. Prominent scoring methods including number right scoring (NR) and negative marking (NM), along with other alternatives, have been studied in an educational system outlining strengths and weaknesses of each method~\cite{Lesage13:Scoring,Scharf07:Assessing}. But approaches composed of diverse modules from available options, i.e., multiple choice, true/false, or explanatory analytical answers lack attention and needs to be evaluated for the effectiveness in assessment of students' abilities.

Effective learning models take into account students' skills and balance the evaluation process accordingly. Question composition and establishment of difficulty levels by dynamic adjustment for scoring has been demonstrated in different learning systems to strengthen the adaptiveness. Prior works have proposed different approaches for student modeling and student motivation, considering the effectiveness of task difficulty and measuring the engagement level to better design adaptive educational systems~\cite{Papouvsek15:Impact}. The inverted U-hypothesis depicts that increases in difficulty should generally pave the way for increase in enjoyment up to some peak point, and afterwards further increases in difficulty lead to decreases in enjoyment. To study the relation between difficulty and enjoyment, Abuhamdeh conducted several experiments to examine and support the findings~\cite{Abuhamdeh12:Importance}. Learning performance curves also have been exploited in studies for adaptive model design~\cite{Martin11:Evaluating}. Generative models that explicitly capture the pairwise knowledge component (skills, procedures, concepts, or facts) relationships to produce a better fit structure reflecting subdivisions in item-type domains with the help of learning curves has been studied~\cite{Pavlik09:Learning}. Another model proposed a modified educational data mining system so that it attains the ability to infer individual student's knowledge component in an adaptive manner~\cite{Pavlik09:Performance}.

Designing an evaluation process which best reflects the proper assessment of each student's ability or effort is a crucial part of every course design. Item Response Theory, based on course structure and appropriate topics to select the k-best questions with adequate difficulty for a particular learner to attain adaptiveness, was brought into focus by Barla~\cite{Barla10:Impact}. Stackelberg game theoretic model was also applied to select  effective and randomized test questions~\cite{Li13:Game} for large scale, public exams  i.e (driver's license test, Toefl iBT). This model  chooses from a predefined set of questions according to the ability level of test taker to compute the optimal test strategies when confidentiality is a concern. Also analysis have been conducted to measure the effects of  grouping student's ability level and achievement using empirical observations~\cite{Burks94:Ability}. Intelligent tutoring system like Cognitive tutors~\cite{Cen07:Over} and REDEEM authoring environment~\cite{Ainsworth04:Evaluating} model assess students' knowledge at different steps and allow teachers to design curricula according to individual skill levels. It has been demonstrated that students' ability or skill inclusion as a parameter resulted in improved accuracy of further prediction to fit observations~\cite{Lee12:Impact}. 

\section{Model}
\label{se:model}
We assume that there are $n$ students and for each student $i$ and for each exam question $j$ there are $m$ real-numbered scores $s_{ij}$ for student $i$'s performance on question $j$. For each student we assume we have a real number $a_i$ that denotes (an approximation of) their ``true ability.'' Ideally, the goal of the exam is to provide as accurate an assessment of the students' true abilities as possible. We seek to find the optimal way in which the exam question could have been weighted in order to give scores for the exam that are as close as possible to the true abilities $a_i$. That is, we seek to obtain weights $w_j$ in order to minimize 
$$\sum_{i=1}^n \left[ a_i - \left(\sum_{j=0}^m \left( w_j \cdot s_{ij} \right) \right) \right]^2,$$
which denotes the mean-squared error between the weighted exam score and the true ability. Note that we can allow a ``dummy'' question 0 with scores of $s_{i,0} = 1$ for all $i$ in order to allow for a constant term with weight $w_0$, which is useful for regression algorithms.
  
\section{Experiments}
\label{se:experiments}
\begin{figure}[thb]
  \begin{center}
  \subfloat[Final grades count]{\includegraphics[width=0.5\textwidth]{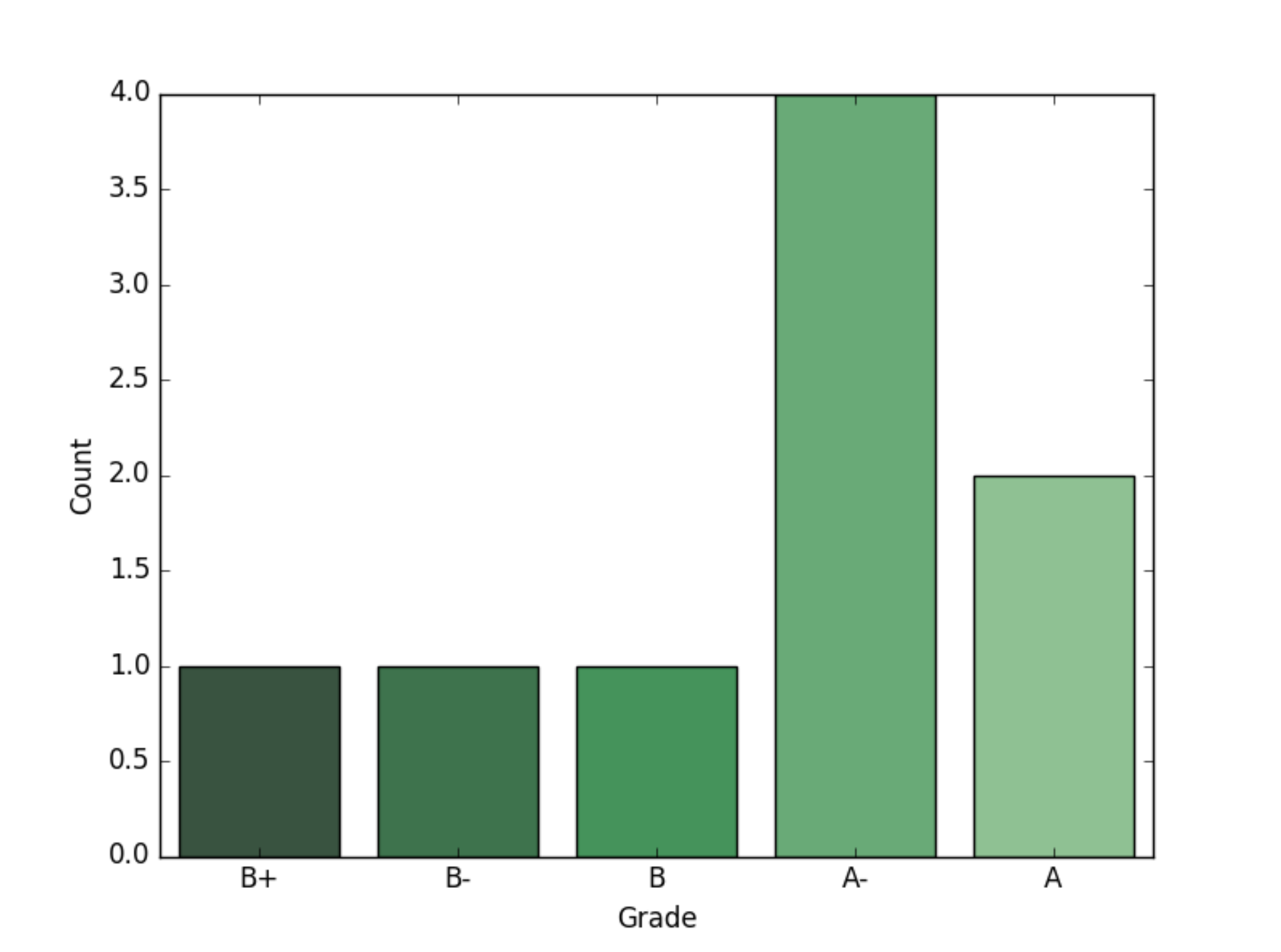}\label{fig:f1}}
  \hfill
  \subfloat[Overall score distribution]{\includegraphics[width=0.5\textwidth]{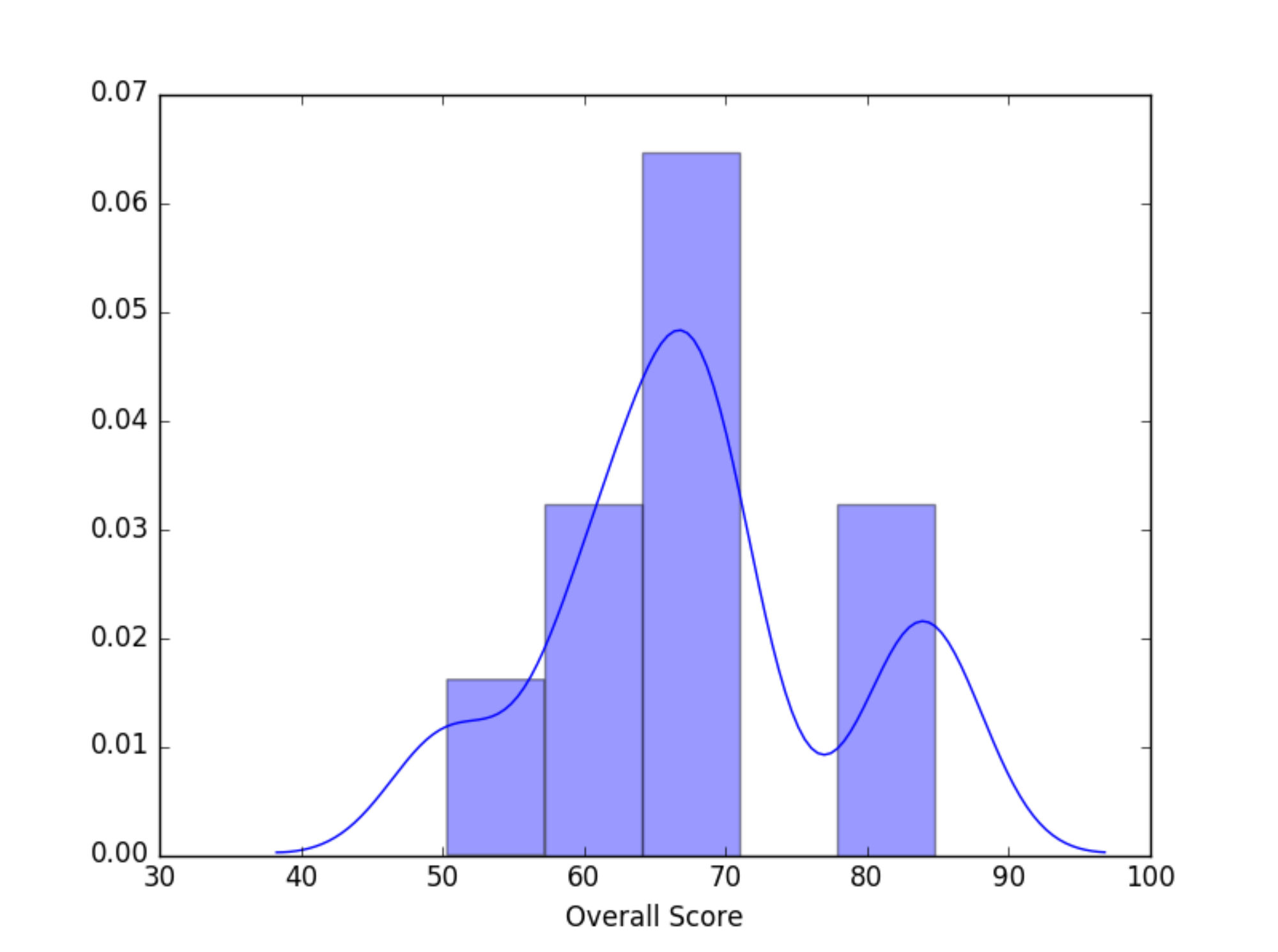}\label{fig:f2}}
\end{center}
  \caption{{\small Overview of student's performance}}
  \label{fig:distributions}
\end{figure}

We experimented on a dataset of graduate level class taught during Spring'17 consisting of nine students. The course curriculum was designed using four major components: Homeworks, Midterm exam, Project, and Final exam with equal percentage (25\% for each) contributing towards the final overall scoring of the students. Final grades were assigned using this overall score as a measure of performance. Though midterm and final exam were considered as two different components, both exams cumulatively contributed to half the score. Average overall score of the students was 67.92 with a standard deviation 10.18. Though the dataset is relatively small, it contains a large degree of variance between students' abilities, and is therefore still representative of interesting phenomena. Distributions of grades and scores are presented in Figure~\ref{fig:distributions}.

%\begin{figure}[thb]
%\begin{center}
%     \includegraphics[width=2.84in]{score_distributions.eps}   
%    
%    {\small Text below figure}         
%\end{center}
%\caption{{\small Caption for Figure.}}
%\end{figure}

%\begin{figure}[thb]
%\begin{center}
%    \includegraphics[width=2.84in]{figure.eps}
%    
%    {\small Text below figure}         
%\end{center}
%\caption{{\small Caption for figure when the caption gets longer than a 
%    single line 
%    and centering really doesn't look so very great at all.}}
%\end{figure}
The final exam was designed with 30 multiple choice question, 15 True/False questions and 5 analytical question which are worth 3, 4 and 10 points each respectively. Each analytical question had several sub parts which resulted in total 53 questions. The average score was 64.27 with standard deviation 27.19. The midterm average was 49.5, which is much lower than the overall score average, and the standard deviation was 19.43. The midterm exam consisted of 30 multiple choice, 15 T/F, and 5 analytical questions with sub parts resulting in a total of 56 questions. For space we omit figures for the midterm, though qualitatively the results were fairly similar to those for the final exam.

Both Final and Midterm exams had average scores that differed from the final overall scores of the students. As a result, to calculate the abilities of the student in the exam we normalized the overall abilities to conform to the final average. Both actual and normalized overall score were used in regression analysis to compare different possibilities. Closed form least squares was implemented to predict the weights of each questions as benchmark. Other approaches involve models i.e., Linear Regression with intercept, Huber regression, and non-negative least squares using variants of objective functions and constraints in  regression analysis. All of these models exploit optimization as a tool to minimize the square loss and approximate the prediction. 

Closed form of ordinary least squares, denoted as normal equation, fits a line passing through the origin, whereas linear model with y-axis intercept do not force the line to pass through the origin, which increases the model capabilities. Huber regression checks outliers' impact on the weights whereas non-negative least squares enforces non-negative constraints for coefficients. Both of the exams were evaluated individually as we are interested in each question even though the pattern of the exam was quite similar. The overall score was normalized using the exam average to represent the exam score and then the coefficient of each question was measured using different approaches to see the extent to which it contributed in the final prediction of students' scores.

\subsection{\bf Closed form Ordinary least squares}  
\begin{figure*}
\begin{minipage}[!hbt]{.45\textwidth}
\centering
\includegraphics[width=0.9\linewidth]{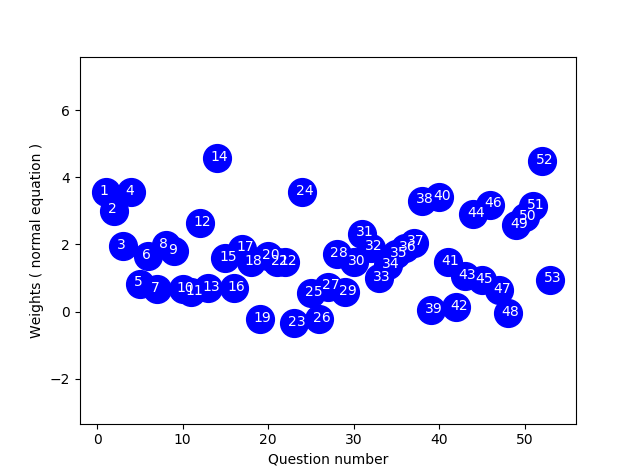}
\caption{\footnotesize Question weights predicting overall score with Closed-Form OLS, for Final Exam}
\label{fi:overall_final_normal}
\end{minipage}\qquad
\begin{minipage}[!hbt]{.45\textwidth}
\centering
\includegraphics[width=0.9\linewidth]{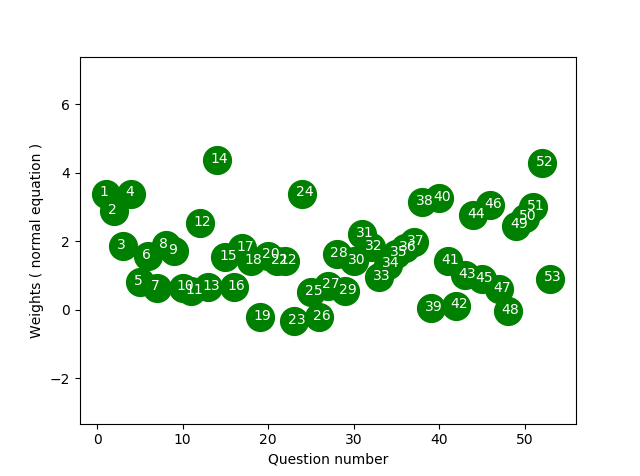}
\caption{\footnotesize Question weights predicting normalized overall score with Closed-Form OLS, for Final Exam}
\label{fi:normalized_final_normal}
\end{minipage}
\end{figure*}

Ordinary least squares (OLS) or linear least squares attempts to estimate the unknown parameters depending on independent explanatory variables. The main objective function is to minimize the sum of the squares of the differences between observed value in the given sample and those predicted by a linear function of a set of features. A closed form solution in linear regression is %the least square equation
$\beta ={(A^{T}A)}^{-1}A^Ty$ where  A is the independent explanatory feature values and y is the observed response or target value. There might be cases where $A^{T}A$ is singular making it non-invertible, so we used the pseudoinverse to solve the equation in our implementation. In general the pseudoinverse is used to solve a system of linear equations where it facilitates the process to compute a best fit solution that lacks a unique solution or to find the minimum norm solution when multiple solutions exist. Figures~\ref{fi:overall_final_normal} and~\ref{fi:normalized_final_normal} show the weights of each final exam question using the ordinary least square (closed-form) approach to fit the actual and normalized overall score. Since the overall score average was close to the final exam average, the weights don't show much deviation with two different scales.

\subsection{Linear Regression with intercept}
Regression through the origin enforces the y-intercept term to be zero in linear models and is used when the line is expected to pass through origin. Linear regression attempts to describe the relationship between a scalar dependent variable y and one or more explanatory variables, i.e., independent variables denoted X.  One of the possibilities for linear regression is to fit a line through the origin and another is to approximate the intercept term so it passes through the center of the datapoints. We used the linear regression approach from scikit-learn python library which uses lapack library from \url{www.netlib.org} to solve the least-squares problem
\begin{displaymath}
\mathop{\mbox{minimize }}_{x} \Vert y - A x {\Vert}_2
\end{displaymath}
\\
Although the objective functions are same, this approach produced a different weighting scheme in comparison to the closed-form method. The linear regression approach from scikit-learn offers two options for approximation to fit the model, one with an intercept term in the equation enhancing the model capability when the line doesn't pass through the origin and the other without such a term. Without the intercept, the linear regression conforms to the parameters found from OLS. Even though an bias term $x_0$ with column vector of all one is introduced in OLS, it favors line passing closer to the origin since the input features are normalized. Whereas with an additional fit\_intercept parameter, the regressor tries to best fit the y-intercept resulting in a better fitted line. Better approximation with intercept can be explained by the target value which is an aggregated score of different components, i.e., homeworks and projects, along with the exam. Since our main goal corresponds to designing an exam which best reflects students' overall ability, the exam scores alone can't represent the expected outcome and sometimes overall score can introduce slightly different observation as they are weighted sum of different components. As a result, linear regression with fit\_intercept seems to be more accurate choice for this experiment and follows the final observation .

Also we have only 8 rows in datasets and more than 50 questions and the problem solved by lapack library takes into consideration the dimension of the matrix of linear equations.  In case of the number of rows being much less than the number of features and rank of A equals to number of rows, there are an infinite number of solutions x which exactly satisfy the equation $y-Ax=0$. Lapack library attempts to find the unique solution of x which minimizes $|x|_2$, and the problem is referred to as finding a minimum norm solution to an underdetermined system of linear equations. Depending on the implementation of the pseudoinverse calculation, the two approaches can result in different optimal weights. 
\begin{figure*}
\centering
\begin{minipage}[!ht]{.45\textwidth}
\centering
\includegraphics[width=0.9\linewidth]{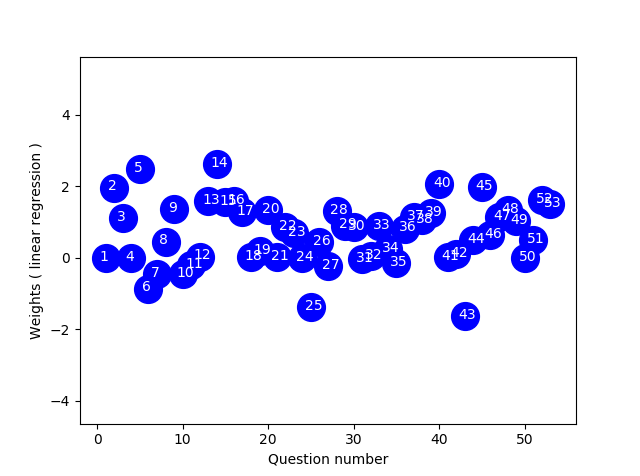}
\caption{\footnotesize Question weights predicting overall score with Linear regression, Final Exam}
\label{fi:overall_final_linear}
\end{minipage}\qquad
\begin{minipage}[!htb]{.45\textwidth}
\centering
\includegraphics[width=0.9\linewidth]{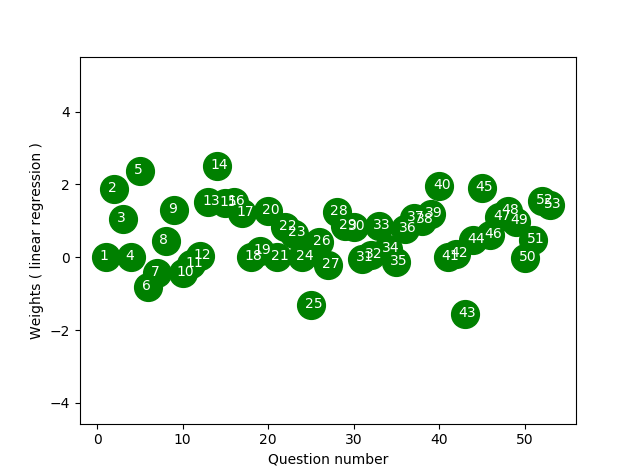}
\caption{\footnotesize Question weights predicting normalized overall score with Linear regression, Final Exam}
\label{fi:normalized_final_linear}

\end{minipage}
\end{figure*}

\begin{comment}
\begin{figure*}
\centering
\begin{minipage}[!ht]{.45\textwidth}
%\includegraphics[height=3in, width=3.4in]{images/midterm_overall_weights_linear.png}
\includegraphics[width=0.9\linewidth]{midterm-overall-weights-linear}
\caption{\footnotesize Question weights predicting overall score with Linear regression, Midterm Exam}
\label{fi:overall_midterm_linear}
\end{minipage}\qquad
\begin{minipage}[!htb]{.45\textwidth}
%\includegraphics[height=3in, width=3.4in]{midterm_normalized_weights_linear.png}
\includegraphics[width=0.9\linewidth]{midterm-normalized-weights-linear}\label{fi:normalized_midterm_linear}

\end{minipage}
\end{figure*}
\end{comment}

\subsection{Huber regression}
Huber regression, which is more robust to outliers, is another linear regression model which optimizes the squared loss for the samples where  $|(y - A'x) / \sigma| < \epsilon $ and the absolute loss for the samples where $|(y - A'x) / \sigma| > \epsilon$, where x and $\sigma$ are parameters to be optimized, y being the target value and $A'x$ is the predicted score. The regularization parameter $\sigma$ ensures that rescaling of y up to certain factor does not affect $\epsilon$ to obtain the same robustness. This method also confirms that the loss function is not as much influenced by the outliers as other samples, while not totally ignoring their effects in the model. In our experiment, we used cross validation to find out the optimal value of $\sigma= 0.1, \epsilon=1.8 $. To control the number of outliers in the sample, $\epsilon$ is used where smaller value of $\epsilon$ ensures more robustness to outliers.
\begin{figure*}
\centering
\begin{minipage}[!ht]{.45\textwidth}
\includegraphics[width=0.9\linewidth]{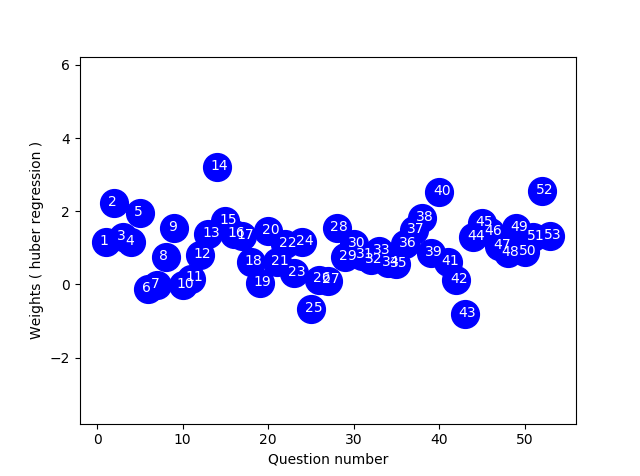}
\caption{\footnotesize Question weights predicting overall score with Huber regression, Final Exam}
\label{fi:overall_final_huber}
\end{minipage}\qquad
\begin{minipage}[!htb]{.45\textwidth}
\centering
\includegraphics[width=0.9\linewidth]{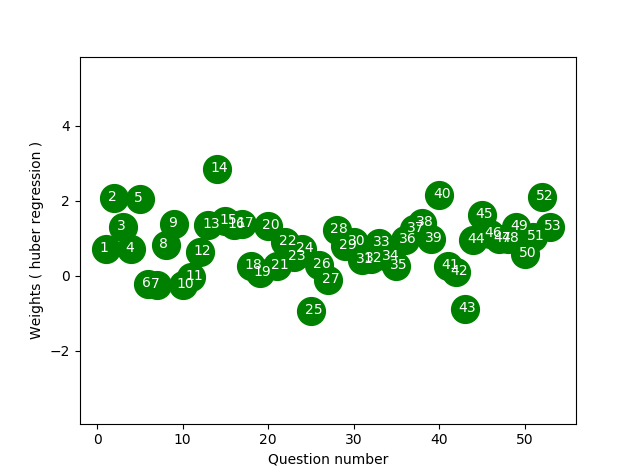}
\caption{\footnotesize Question weights predicting normalized overall score with Huber regression, Final Exam}
\label{fi:normalized_final_huber}

\end{minipage}
\end{figure*}

\begin{comment}
\begin{figure*}
\centering
\begin{minipage}[!ht]{.45\textwidth}

%\includegraphics[height=3in, width=3.4in]{midterm_overall_weights_huber.png}
\includegraphics[width=0.9\linewidth]{midterm-overall-weights-huber}
\caption{Question weights predicting overall score with Huber regression, Midterm Exam}
\label{fi:overall_midterm_huber}
\end{minipage}\qquad
\begin{minipage}[!htb]{.45\textwidth}
%\includegraphics[height=3in, width=3.4in]{midterm_normalized_weights_huber.png}
\includegraphics[width=0.9\linewidth]{midterm-normalized-weights-huber}
\caption{\small Question weights predicting normalized overall score with Huber regression, Midterm Exam}
\label{fi:normalized_midterm_huber}

\end{minipage}
\end{figure*}
\end{comment}

\subsection{Non negative least squares  regression }
\begin{figure*}
\centering
\begin{minipage}[!ht]{.45\textwidth}
\centering
\includegraphics[width=0.9\linewidth]{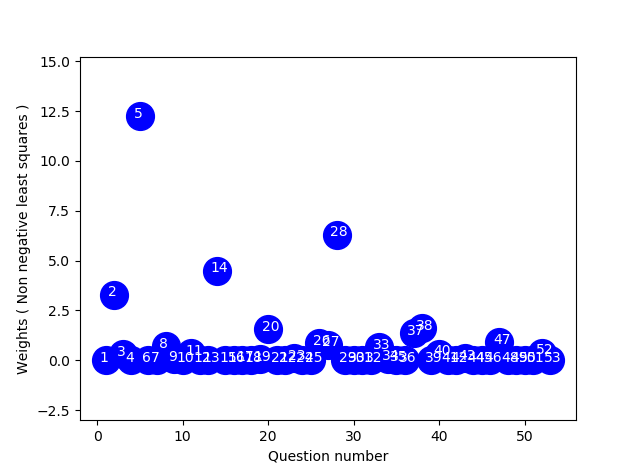}
\caption{\footnotesize Question weights predicting overall score with Non-negative least squares, Final Exam}
\label{fi:overall_final_nnls}
\end{minipage}\qquad
\begin{minipage}[!htb]{.45\textwidth}
\centering
\includegraphics[width=0.9\linewidth]{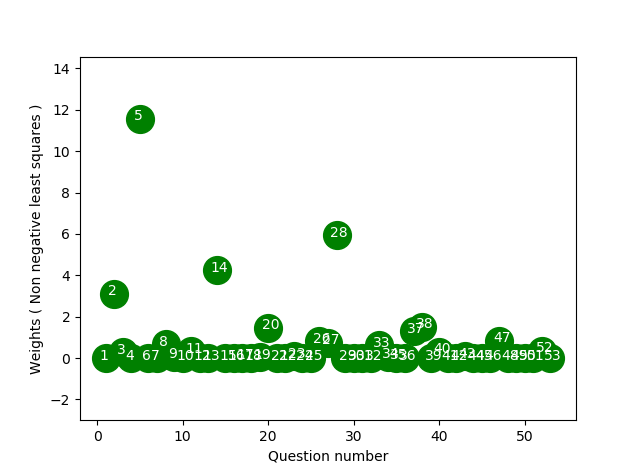}
\caption{\footnotesize Question weights predicting normalized overall score with Non-negative least squares, Final Exam}
\label{fi:normalized_final_nnls}

\end{minipage}
\end{figure*}

Non-negative least squares (NNLS) is a constrained version of the least squares problem in mathematical optimization where only non-negative coefficients are allowed. That is, given a matrix A and a column vector of response variables y, the goal is to find ${arg\,min}_x  {\|\mathbf{Ax} -\mathbf{y} \|}_{2}$ subject to $x \ge 0$.
Here $x \ge 0$ means that each component of the vector x should be non-negative. As we are interested in designing an exam, approaches defining constraints with only positive weights question can be effective, since it seems unnatural to assign negative weight to an exam question. Figure\ref{fi:overall_final_nnls} and Figure~\ref{fi:normalized_final_nnls} shows the non-zero weights for the final exam questions. NNLS method from scipy library was used to solve the constrained optimization problem to calculate the weights for this purpose.

\begin{comment}
\begin{figure*}
\centering
\begin{minipage}[!ht]{.45\textwidth}

\includegraphics[height=3in, width=3.4in]{midterm-overall-weights-nnls}
\caption{\small Question weights predicting overall score with Non-negative least squares, Midterm Exam}
\label{fi:overall_midterm_nnls}
\end{minipage}\qquad
\begin{minipage}[!htb]{.45\textwidth}
\includegraphics[height=3in, width=3.4in]{midterm-normalized-weights-nnls}
\caption{\scriptsize Question weights predicting normalized overall score with Non-negative least squares, Midterm Exam}
\label{fi:normalized_midterm_nnls}

\end{minipage}
\end{figure*}
\end{comment}

\begin{figure*}
\centering
\begin{minipage}[!ht]{.45\textwidth}

\includegraphics[width=0.9\linewidth]{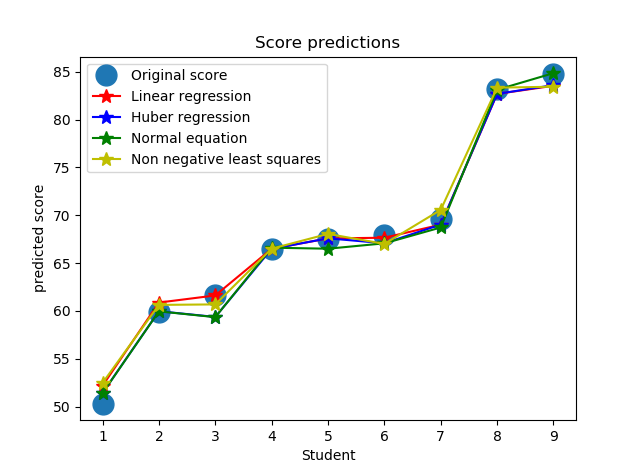}
\caption{\footnotesize Overall score prediction, Final Exam}
\label{fi:overall_final_predictions}
\end{minipage}\qquad
\centering
\begin{minipage}[!htb]{.45\textwidth}

\includegraphics[width=0.9\linewidth]{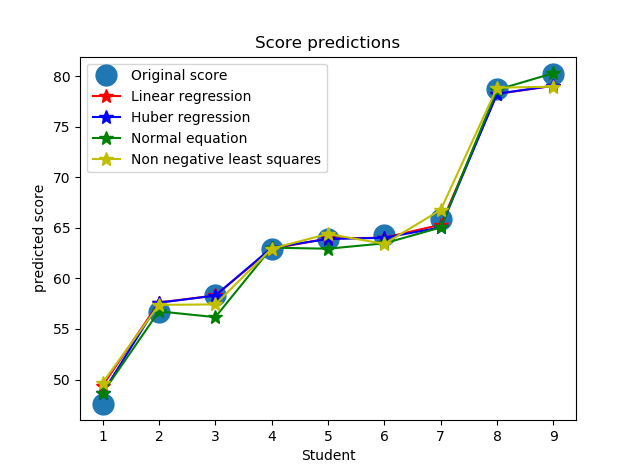}
\caption{\footnotesize Normalized score prediction, Final Exam}
\label{fi:normalized_final_predictions}

\end{minipage}
\end{figure*}

\subsection{Comparing the approaches}
Predicted score for final exam using all four approaches produces low error as shown in Figure~\ref{fi:overall_final_predictions}
and Figure~\ref{fi:normalized_final_predictions} in comparison to uniform weighting and the designed weighting used in the actual exam. As a measurement of performance evaluation of different approaches, mean absolute error is tabulated in Table~\ref{ta:MAE} for both exams with two different scale of score, normalized and actual, respectively. Uniform weighting method where all questions are worth equal amounts is used as an benchmark to compare with the proposed method. Also the actual weighting in the exam, to predict the overall score, is taken into consideration to check how much they conform with the final score. For model formulation, the leave one out cross validation approach was used and finally the weights from all the models were averaged to compute the final questions weights, which are used to predict the overall score. 

\begin{comment}
\begin{figure*}
\centering
\begin{minipage}[!hbt]{.45\textwidth}
\centering
%\includegraphics[height=3in, width=3.4in]{midterm_overall_predictions.png}
\includegraphics[width=0.9\linewidth]{midterm-overall-predictions}
\caption{\small{Overall score prediction, Midterm Exam}}
\label{fi:overall_midterm_predictions}
\end{minipage}\qquad
\begin{minipage}[!hbt]{.45\textwidth}
%\includegraphics[height=3in, width=3.4in]{midterm_normalized_predictions.png}
\includegraphics[width=0.9\linewidth]{midterm-normalized-predictions}
\caption{\small{Normalized overall score prediction, Midterm Exam}}
\label{fi:normalized_midterm_predictions}
\end{minipage}
\end{figure*}
\end{comment}

%Predicted score using all  approaches is demonstrated in Figure ~\ref{fi:overall_midterm_predictions} and ~\ref{fi:normalized_midterm_predictions}. 
%As a measurement of performance evaluation of different approaches, mean absolute error  is tabulated in Table~\ref{ta:MAE} for both exam with two different scale of score, normalized and actual respectively. Uniform weighting method where each question's worth equal amount is used as an benchmark to compare with the proposed method. Also the actual weighting in the exam, to predict the overall score, is taken into consideration to check how much they conform with the final score.
\renewcommand{\tabcolsep}{4pt}
\begin{table*}[!htb]
\scriptsize
 \caption{{\small Comparison of Mean absolute error}}
\begin{center}
\begin{tabular}{|*{7}{c|}} \hline
Overall Score&Uniform weights & Actual Weights &  Linear Regression & Huber Regression & Ordinary LS & Non-Negative LS \\ \hline
Final ( Normalized) &7.2368	& 6.1644 &	0.5690 & 0.5280& 0.6804 & 0.8135\\ \hline
Midterm (Normalized) &2.9898	& 3.2856	&0.3802 & 0.3967 & 0.3161 & 0.5529\\ \hline
Final (Actual) & 8.0209	& 6.3726 &	0.6013 & 0.7234 & 0.7190 & 0.8597\\ \hline
Midterm (Actual) & 17.5650	& 17.51134	& 0.5218 & 0.5094 & 0.4338& 0.7587\\ \hline
\end{tabular}
\end{center}
\label{ta:MAE}
\end{table*}

\section{Discussion}
While the approaches we use are existing techniques for linear regression, we encountered several issues of potentially more general theoretical interest in our setting. 

\subsection{Overall Score Computation}
\begin{figure*}
\centering
\begin{minipage}[!ht]{.45\textwidth}

\includegraphics[width=0.9\linewidth]{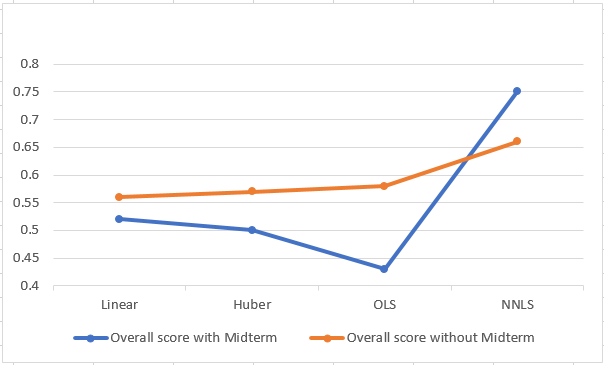}
\caption{\footnotesize Comparison of prediction errors for Midterm and Overall score with and without Midterm}
\label{fi:comparison_prediction_error_midterm}
\end{minipage}\qquad
\centering
\begin{minipage}[!htb]{.45\textwidth}

\includegraphics[width=0.9\linewidth]{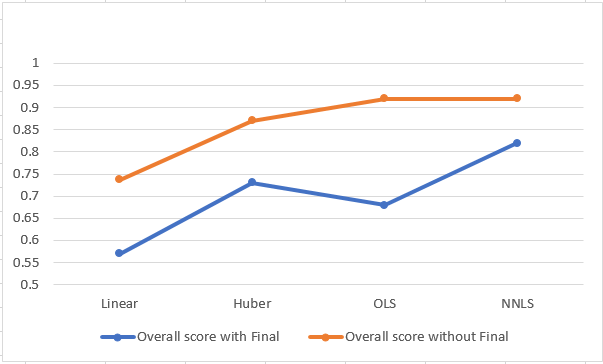}
\caption{\footnotesize Comparison of prediction errors for Final, Overall score with and without Final Score}
\label{fi:comparison_prediction_error_final}

\end{minipage}
\end{figure*}

One of the major concerns was how to incorporate all the components' information for overall score computation.  Since we are using overall score to compute exam questions' weights and overall score already contains that particular exam's weighted score, it should be taken into consideration whether to include it in overall score computation or not. But excluding an component from computation will result in information loss. As a result, we experimented on both approaches of overall score computation to observe the change in weights and prediction errors. From Figure~\ref{fi:comparison_prediction_error_midterm} and Figure~\ref{fi:comparison_prediction_error_final}, it is evident that overall score which includes all components performs better in both exams except the non-negative least square approach for Midterm one. Also we observed that changes in weights due to exclusion of an component are trivial and almost proportional. 

\subsection{Unique vs. multiplicity of solutions for linear regression, depending on the rank of the matrix}
System of linear equations or a system of polynomial equations is referred as underdetermined if number of equations available are less than unknown parameters. Unknowns parameters in a model represents an available degree of freedom, whereas each of the equations puts a constraint restricting the degree of freedom along a particular axis. When the number of equations is less than the number of unknown parameters, the system becomes underdetermined due to available degrees of freedom along some axes. As a result an underdetermined system can have infinitely many solutions or no solution at all. Since in our case study, the system is underdetermined and also A is singular, $A^TA$ is also singular, and the equation$ A^TAx=A^Ty$ had infinitely many solutions. The pseudoinverse is a way to choose a "best solution" $x^+=A^+y$.

\subsection{Intuition for negative weights}
The weights denote the relative contributions in the solution. They shows relative impact in the predictions compared to the other features in the samples. As a result, we can think of the negative weight questions in our sample as less important than the ones with high and positive coefficients. While for many settings it may make perfectly good sense to include negative weights (for example in financial forecasting), it may not be sensible for exam questions, as it would mean that students are incentivized to get those questions wrong.

\subsection{Results for truncating weights at 0, and description of algorithm for doing this}
In order to limit the coefficients of linear equation to be only positive, we used non-negative least squares where the objective function includes an additional assumption that weights are non-negative and then solves the system~\cite{chen09:Nonnegativity}. NNLS from the scipy optimization library solves the system of linear equation with non-negativity constraints which served our purpose in the experiment.

\subsection{Variants of linear regression in python}
The closed-form normal equation uses the dot product and inverse of matrix to solve for unknown parameters in a system of linear equations. A bias parameter with all ones is used to introduce a constant term in the matrix. Since it doesn't take into consideration offset of the line from mean while fitting the intercept, approximation error increases. However linear regression in scikit-learn~\cite{Pedregosa11:Scikit} does not compute the inverse of A. Instead it uses Lapack driver routine xGELS to solve least squares on the assumption that  $\mbox{rank}(A) = \min(m,n)$. xGELS makes use of  QR or LQ factorization of A which can result in different coefficients than the prior discussed methods. Not only that, the fit\_intercept term in scikit-learn represents the Y-intercept of the regression line which is the value predicted when all the independent variables are zero at a time. On top of that, without the intercept term, the model itself become biased and all the other parameters get affect due to the fact that the bias term in OLS is not scaled but only an approximation with all one column vector. Also due to normalization on dataset, when any  question is answered by all the students, the matrix bias term and that particular column become identical and is assigned same weight with OLS. As a result, including the intercept term results in better weights which ouputs the scaled value after the coefficient calculation and produces different results than the one with no intercept.

\subsection{Closer analysis of certain notable questions}
We take a closer look at several notable questions that stood out from the extreme weights they were given in the regression output. First, multiple choice question 25 in the final exam was given a negative weight of -1.307. The full distribution table is given in Table~\ref{ta:MC_low_linear}. We can see that the two students with highest overall score got the question wrong, while several of the weaker students got the question right, which provides an explanation for the negative weight given.

\begin{table}[!hbt]
\scriptsize
 \caption{{\footnotesize Multiple choice 25 score distribution in Final; given lowest in linear regression}}
\begin{center}
\begin{tabular}{|*{3}{c|}} \hline
Student& Obtained score & Final Score \\ \hline
1 & 1 	& 50.32\\ \hline
2  & 0	 & 59.89\\ \hline
3 & 0		& 61.63 \\\hline
4 & 1		& 66.50\\ \hline
5 & 1		 & 67.54\\ \hline
6 & 0 	& 67.92\\ \hline
7 & 1 	& 69.57\\ \hline
8 & 0 	& 83.16\\ \hline
9 & 0	&	 84.73\\ \hline
\end{tabular}
\end{center}
\label{ta:MC_low_linear}
\end{table}

Next, multiple choice question 5 from the final exam had a very large weight of 2.376. Its distribution is given in Table~\ref{ta:MC_high_linear_normalized}. We can see that the strongest three overall students got this question right, while the weakest 6 got it wrong, which justifies the high weight. Similarly, question 6 in which the two strongest students were the only ones to answer correctly also received a very high, but slightly lower, weight of 1.624, as shown in Table~\ref{ta:MC_high_linear}.

\begin{table}[!hbt]
\scriptsize
 \caption{{\footnotesize Multiple choice 5 score distribution in Final; given high weight in linear regression}}
\begin{center}
\begin{tabular}{|*{3}{c|}} \hline
Student& Obtained score & Final Score \\ \hline
1 & 0 	& 50.32\\ \hline
2  & 0	 & 59.89\\ \hline
3 & 0		& 61.63 \\\hline
4 & 0	&	 66.50\\ \hline
5 & 0	 &	 67.54\\ \hline
6 & 0 	& 67.92\\ \hline
7 & 1 	& 69.57\\ \hline
8 &1		& 83.16\\ \hline
9 & 1		&	 84.73\\ \hline
\end{tabular}
\end{center}
\label{ta:MC_high_linear_normalized}
\end{table}

\begin{table}[!hbt]
\scriptsize
 \caption{{\footnotesize Multiple choice 6 score distribution in Midterm normalized; given high weight linear regression}}
\begin{center}
\begin{tabular}{|*{3}{c|}} \hline
Student& Obtained score & Final Score \\ \hline
1 & 0 	& 36.67\\ \hline
2  & 0	 & 43.65\\ \hline
3 & 0		& 44.92 \\\hline
4 & 0	&	 48.46\\ \hline
5 & 0	 &	 49.23\\ \hline
6 & 0 	& 49.50\\ \hline
7 &0 	&  50.70\\ \hline
8 &1		& 60.61\\ \hline
9 & 1		&	 61.75\\ \hline
\end{tabular}
\end{center}
\label{ta:MC_high_linear}
\end{table}

Finally, we can see the table for another question with a negative weight, where generally the weaker students in the class performed better than the stronger students---question 1a from the midterm for normalized overall score, with weight -1.048, given in Table~\ref{ta:MC_low_linear_regression}.

\begin{table}[!hbt]
\scriptsize
 \caption{{\footnotesize Analytical question 1(a) score distribution in Midterm normalized; given low weight in linear regression}}
\begin{center}
\begin{tabular}{|*{3}{c|}} \hline
Student& Obtained score & Normalized Final Score \\ \hline
1 & 0.5 	& 36.67\\ \hline
2  & 1	 & 43.65\\ \hline
3 & 1		& 44.92 \\\hline
4 & 0.75	&	 48.46\\ \hline
5 & 1	 &	 49.23\\ \hline
6 & 0.75 	& 49.50\\ \hline
7 &0 	&  50.70\\ \hline
8 & 0		& 60.61\\ \hline
9 & 0		&	 61.75\\ \hline
\end{tabular}
\end{center}
\label{ta:MC_low_linear_regression}
\end{table}

\subsection{Effect of normalization on the approaches}
We explored how it would effect the results to divide all approaches by their mean before/after applying them. Final overall score is the factored aggregation of different components constituent of homeworks, two exams, and project. But normalizing the overall score with their respective exam mean ratio results in relatively better outcome since we are trying to relate the exam weights with their normalized ability. In the final exam, the class average did not deviate much from overall average, so the mean absolute error difference for Actual and Normalized approaches are very low. In the contrary, in midterm average score of the class was 49.5 which is much lower than the overall average of 67.92. So multiplying the overall score by the midterm mean ratio resulted in more precise prediction for this case.

\subsection{Additional observations}
%--Different approaches for all 0 or all 1
We examined the results of the approaches for a question that all students answered correctly (multiple choice 1), with results given by Table~\ref{ta:everyone_answered_midterm}. The weights were zero when none of the students answered a question correctly irrespective of the approaches. In the final when only the highest scorer answered correctly, different approaches demonstrated variations in their weighting, i.e, the linear approach put much higher weight in comparison to the other approaches (Table~\ref{ta:hs_answered_final}).

%--Duplicate questions how will they be weighted 
In final MC 1 and MC 4 both were answered correctly by all the students (and therefore can be viewed as a ``duplicate'' question). All the approaches except NNLS weighted the questions similarly, as shown by Table~\ref{ta:duplicate_answered_final}.	

\begin{table}[!hbt]
\scriptsize
 \caption{{\footnotesize Comparison of weights for different approaches. Everyone answered MC1 correctly.}}
\begin{center}
\begin{tabular}{|*{3}{c|}} \hline
Approach & Overall  & Normalized  \\ \hline
Linear &	0		&	0  \\ \hline
Huber &	0.4098 &	0.2610  \\ \hline
OLS &	3.6673  &	2.6725   \\ \hline
NNLS&	0 &	35.4471  \\ \hline
\end{tabular}
\end{center}
\label{ta:everyone_answered_midterm}
\end{table}

\begin{table}[!hbt]
\scriptsize
 \caption{{\footnotesize Comparison of weights for different approaches. Only highest scorer answered correctly AE 1(c).}}
\begin{center}
\begin{tabular}{|*{3}{c|}} \hline
Approach & Overall  & Normalized \\ \hline
Linear &	1.3431 &	1.2710 \\ \hline
Huber &	0.7163 &	0.7295\\ \hline
OLS &	 	-0.0501	&-0.0474  \\ \hline
NNLS&	0 & 0 \\ \hline
\end{tabular}
\end{center}
\label{ta:hs_answered_final}
\end{table}

\begin{table}[!hbt]
\scriptsize
 \caption{{\footnotesize Comparison of weights for MC1 and MC4, which were answered correctly by everyone in the final}}
\begin{center}
\begin{tabular}{|*{5}{c|}} \hline
\multirow{2}{*}{Approach}  &\multicolumn{2}{c|}{MC 1}&\multicolumn{2}{c|}{MC 4}\\
\cline{2-5}
  & Overall  & Normalized&Overall  & Normalized \\ \hline
Linear &	1.97E-16 &	0		&			2.47E-17	&0 \\ \hline
Huber &	1.4977&	1.2033	&	1.4977	&1.2033\\ \hline
OLS &	 	3.5467	&	3.3563	&	3.5467&	3.3563 \\ \hline
NNLS&	0&0	&	0 &	0\\ \hline
\end{tabular}
\end{center}
\label{ta:duplicate_answered_final}
\end{table}

%-- Argument for diversity of questions as opposed to just one ``great'' question

%Another major concern for course design is to keep up the student's engagement and motivation throughout. As stated above in~\ref{se:intro}, setting the difficulty level  closely relates to engagement according to inverted-U hypothesis. As a result, to effectively  identify  the ability of all students with rigorous evaluation process supports in favor of  diversity  of questions, components and methods to increase the engagement and motivation rather than just selecting one best question/methods which can help distinguish the student's performance.

\section{Conclusion}
\label{se:conc}
The approaches demonstrate that, at least according to our model, novel exam question weighting policies could lead to significantly better assessments of students' performance. We showed that our approaches lead to significantly lower mean squared error when optimizing weights on midterm and final exam questions in order to most closely approximate the overall final score, which we view as a proxy for the true student's ability. From analyzing the optimal weights we identified several questions that stood out, and have a better understanding of what it means for a question to be ``good'' and ``bad.'' We also described several practical factors that would need to be taken into consideration for application of our approaches to real examinations.
%\clearpage

% some example references generated by using \bibliographystyle{natbib}
%\setlength{\bibhang}{0.75cm}
%\begin{small}
\bibliographystyle{plain}
\bibliography{D://FromBackup/Research/refs/dairefs}

\begin{thebibliography}{10}

\bibitem{Abuhamdeh12:Importance}
Sami Abuhamdeh and Mihaly Csikszentmihalyi.
\newblock The importance of challenge for the enjoyment of intrinsically
  motivated, goal-directed activities.
\newblock {\em Personality and Social Psychology Bulletin}, 38(3):317--330,
  2012.

\bibitem{Ainsworth04:Evaluating}
Shaaron Ainsworth and Shirley Grimshaw.
\newblock Evaluating the redeem authoring tool: can teachers create effective
  learning environments?
\newblock {\em International Journal of Artificial Intelligence in Education},
  14(3, 4):279--312, 2004.

\bibitem{Barla10:Impact}
Michal Barla, M{\'a}ria Bielikov{\'a}, Anna~Bou Ezzeddinne, Tom{\'a}{\v{s}}
  Kram{\'a}r, Mari{\'a}n {\v{S}}imko, and Oto Voz{\'a}r.
\newblock On the impact of adaptive test question selection for learning
  efficiency.
\newblock {\em Computers \& Education}, 55(2):846--857, 2010.

\bibitem{Burks94:Ability}
L~Burks.
\newblock Ability group level and achievement.
\newblock {\em School Community Journal}, 4(1):11--24, 1994.

\bibitem{Cen07:Over}
Hao Cen, Kenneth~R Koedinger, and Brian Junker.
\newblock Is over practice necessary?-improving learning efficiency with the
  cognitive tutor through educational data mining.
\newblock {\em Frontiers in Artificial Intelligence and Applications}, 158:511,
  2007.

\bibitem{chen09:Nonnegativity}
Donghui Chen and Robert~J Plemmons.
\newblock Nonnegativity constraints in numerical analysis.
\newblock {\em The birth of numerical analysis}, 10:109--140, 2009.

\bibitem{Kirkpatrick11:Negative}
Robert Kirkpatrick and Yuebing Zang.
\newblock The negative influences of exam-oriented education on chinese high
  school students: Backwash from classroom to child.
\newblock {\em Language Testing in Asia}, 1(3):36, 2011.

\bibitem{Lee12:Impact}
Jung~In Lee and Emma Brunskill.
\newblock The impact on individualizing student models on necessary practice
  opportunities.
\newblock {\em International Educational Data Mining Society}, 2012.

\bibitem{Lesage13:Scoring}
Ellen Lesage, Martin Valcke, and Elien Sabbe.
\newblock Scoring methods for multiple choice assessment in higher
  education--is it still a matter of number right scoring or negative marking?
\newblock {\em Studies in Educational Evaluation}, 39(3):188--193, 2013.

\bibitem{Li13:Game}
Yuqian Li and Vincent Conitzer.
\newblock Game-theoretic question selection for tests.
\newblock In {\em IJCAI}, pages 254--262, 2013.

\bibitem{Martin11:Evaluating}
Brent Martin, Antonija Mitrovic, Kenneth~R Koedinger, and Santosh Mathan.
\newblock Evaluating and improving adaptive educational systems with learning
  curves.
\newblock {\em User Modeling and User-Adapted Interaction}, 21(3):249--283,
  2011.

\bibitem{Norcini03:Setting}
John~J Norcini.
\newblock Setting standards on educational tests.
\newblock {\em Medical education}, 37(5):464--469, 2003.

\bibitem{Papouvsek15:Impact}
Jan Papou{\v{s}}ek and Radek Pel{\'a}nek.
\newblock Impact of adaptive educational system behaviour on student
  motivation.
\newblock In {\em International Conference on Artificial Intelligence in
  Education}, pages 348--357. Springer, 2015.

\bibitem{Pavlik09:Learning}
Philip~I Pavlik~Jr, Hao Cen, and Kenneth~R Koedinger.
\newblock Learning factors transfer analysis: Using learning curve analysis to
  automatically generate domain models.
\newblock In {\em Educational Data Mining 2009}, 2009.

\bibitem{Pavlik09:Performance}
Philip~I Pavlik~Jr, Hao Cen, and Kenneth~R Koedinger.
\newblock Performance factors analysis--a new alternative to knowledge tracing.
\newblock {\em Online Submission}, 2009.

\bibitem{Pedregosa11:Scikit}
F.~Pedregosa, G.~Varoquaux, A.~Gramfort, V.~Michel, B.~Thirion, O.~Grisel,
  M.~Blondel, P.~Prettenhofer, R.~Weiss, V.~Dubourg, J.~Vanderplas, A.~Passos,
  D.~Cournapeau, M.~Brucher, M.~Perrot, and E.~Duchesnay.
\newblock Scikit-learn: Machine learning in {P}ython.
\newblock {\em Journal of Machine Learning Research}, 12:2825--2830, 2011.

\bibitem{Scharf07:Assessing}
Eric~M Scharf and Lynne~P Baldwin.
\newblock Assessing multiple choice question (mcq) tests-a mathematical
  perspective.
\newblock {\em Active Learning in Higher Education}, 8(1):31--47, 2007.

\bibitem{Scouller98:Influence}
Karen Scouller.
\newblock The influence of assessment method on students' learning approaches:
  Multiple choice question examination versus assignment essay.
\newblock {\em Higher Education}, 35(4):453--472, 1998.

\end{thebibliography}
%\bibliography{question_weighting}
%\end{small}

\end{document}